\documentclass[8.5pt,twoside,twocolumn]{article}
\usepackage[super,sort&compress,comma]{natbib} 
\usepackage{mhchem}
\usepackage{times,mathptmx}
\usepackage{times}
\usepackage{sectsty}
\usepackage{balance} 

\usepackage{graphicx} 
\usepackage{lastpage}
\usepackage[format=plain,justification=raggedright,singlelinecheck=false,font=small,labelfont=bf,labelsep=space]{caption} 
\usepackage{fancyhdr}
\begin{document}

\thispagestyle{plain}
\fancypagestyle{plain}{
\renewcommand{\headrulewidth}{1pt}}
\renewcommand{\thefootnote}{\fnsymbol{footnote}}
\renewcommand\footnoterule{\vspace*{1pt}%
\hrule width 3.4in height 0.4pt \vspace*{5pt}} 
\setcounter{secnumdepth}{5}

\makeatletter 
\def\subsubsection{\@startsection{subsubsection}{3}{10pt}{-1.25ex plus -1ex minus -.1ex}{0ex plus 0ex}{\normalsize\bf}} 
\def\paragraph{\@startsection{paragraph}{4}{10pt}{-1.25ex plus -1ex minus -.1ex}{0ex plus 0ex}{\normalsize\textit}} 
\renewcommand\@biblabel[1]{#1}            
\renewcommand\@makefntext[1]%
{\noindent\makebox[0pt][r]{\@thefnmark\,}#1}
\makeatother 
\renewcommand{\figurename}{\small{Fig.}~}
\sectionfont{\large}
\subsectionfont{\normalsize} 

\fancyfoot{}
\fancyfoot[R]{\footnotesize{\sffamily{1--\pageref{LastPage} ~\textbar  \hspace{2pt}\thepage}}}
\fancyhead{}
\renewcommand{\headrulewidth}{1pt} 
\renewcommand{\footrulewidth}{1pt}
\setlength{\arrayrulewidth}{1pt}
\setlength{\columnsep}{6.5mm}
\setlength\bibsep{1pt}

\twocolumn[
  \begin{@twocolumnfalse}
\noindent\textit{\small{\textbf{Working draft}}}  \newline
\noindent\Large{\textbf{A minimal description of morphological hierarchy in two-dimensional aggregates}}
\vspace{0.6cm}

\noindent\large{Tamoghna Das,\textit{$^{a}$} T. Lookman,\textit{$^{b}$} and M. M. Bandi$^{\ast}$\textit{$^{c}$}}



\vspace{0.6cm}

\noindent\small{
A dimensionless parameter $\Lambda$ is proposed to describe a hierarchy of morphologies in two-dimensional ($2D$) aggregates formed due to varying competition between short-range attraction and long-range repulsion. Structural transitions from finite non-compact to compact to percolated structures are observed in the configurations simulated by molecular dynamics at a constant temperature and density. Configurational randomness across the transition, measured by the two-body excess entropy $S_2$, exhibits data collapse with the average potential energy $\bar{\mathcal{E}}$ of the systems. Independent master curves are presented among $S_2$, the reduced second virial coefficient $B_2^*$ and $\Lambda$, justifying this minimal description. This work lays out a coherent basis for the study of 2D aggregate morphologies relevant to diverse nano- and bio-processes.
}
\vspace{0.7cm}
 \end{@twocolumnfalse}
  ]

\footnotetext{\textit{$^{a}$~ Collective Interactions Unit, OIST Graduate University, Onna, Okinawa 9040495, Japan. E-mail: tamoghna.das@oist.jp}}
\footnotetext{\textit{$^{b}$~ Theoretical Division, Los Alamos National Laboratory, Los Alamos, NM 87545, USA. E-mail: txl@lanl.gov}}
\footnotetext{\textit{$^{c}$~ Collective Interactions Unit, OIST Graduate University, Onna, Okinawa 9040495, Japan. Corresponding author: bandi@oist.jp}}

\section{Introduction}
Aggregates, due to their finite spatial correlations, are structurally intermediate between completely random (liquid) and ordered (crystalline) states. Formation of these aggregates requires competition \cite{lebo_pen,lebo_pen2} between very short range attraction and long-range repulsion together with suitable thermodynamic conditions (low temperature and density). The structure of aggregates may assume different shapes with varied degrees of randomness controlled by the competition \cite{modphs}. This self-assembly of particles is generic and has been observed in systems including polymer-coated colloids \cite{colloid_crumb, colloid_crumb2}, globular proteins in weakly polar solvents \cite{protein_soup,protein_soup2,protein_soup3,protein_soup4}, quantum dots and nano-particles \cite{more_cooking,more_cooking2,more_cooking3}. In addition, the competition can be tuned in several ways for a specific system which is advantageous from the perspective of industrial applications. For example, the range of attraction in a colloidal system can be modified by suitable chemical treatment or choice of polymer coating, whereas changing the salt concentration or pH balance of solvent controls the effect of repulsion in the same system \cite{salt}. The diversity of systems and the flexibility of control have, however, barred a generic account of aggregate morphologies required for further progress in developing nano- and bio-metric functional materials \cite{new_cookies,new_cookies2,new_cookies3,new_cookies4,new_cookies5,new_cookies6}. Here we propose a minimal description based on a single dimensionless parameter.

The descriptor $\Lambda$ is a ratio of two effective lengths computed from the competing parts of pair interactions. Tuning the interactions, a set of configurations is simulated by using molecular dynamics. A structural hierarchy within the generated configurations is then characterised by employing standard statistical diagnostics. Next, the degree of positional randomness governing the hierarchy is quantified by two-body excess entropy $S_2$. This enables us to describe all observed morphologies in terms of the average potential energy $\bar{\mathcal{E}}$ and $\Lambda$, independently. Collapse of the reduced second virial $B_2^*$, computed for all morphologies, onto a master curve as a function of $\Lambda$ provides experimental access to this description. The organisation of the rest of the paper is as follows. In section 2 we introduce the competing interactions and their control parameters including the morphological variations that result from tuning the competition with a brief description of simulation details. The spatial correlations for the simulated configurations and several statistical characterisations of the same are presented in section 3. The descriptor $\Lambda$ is introduced in section 4. In section 5, we quantify the structural randomness of observed aggregates and relate it with their respective mean energy. We discuss further implications of our findings in sec-tion 6 followed by a brief concluding summary.

\section{Competing interactions and variations in morphology}
\begin{figure*}
\centering
\includegraphics[width=0.85\linewidth]{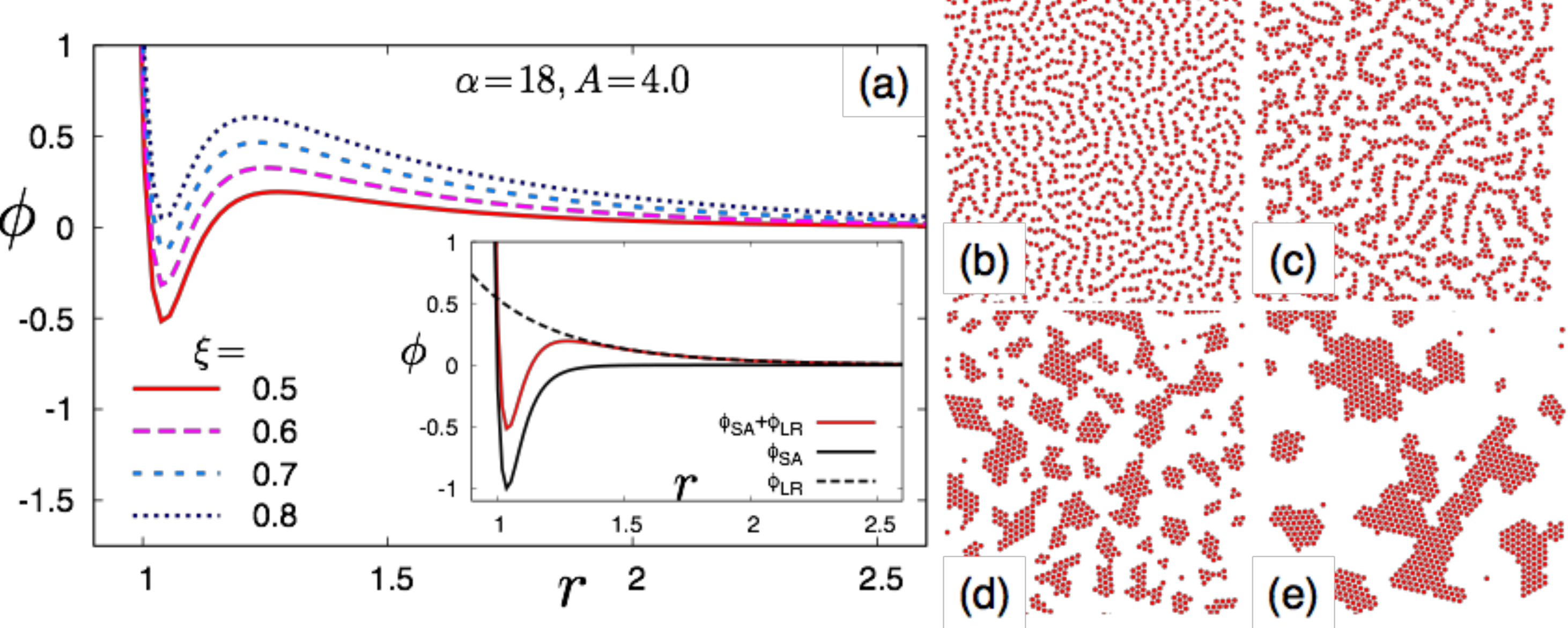}
\caption{
(color online) (a) Variation of the effective potential profile $\phi$ (defined in text) with $\xi$ is plotted for $A\!=\!4.0$ and $\alpha\!=\!18$. Note that the attractive minimum goes from global to local as a function of increasing $\xi$. {\it Inset} shows the attractive $\phi_{SA}$ and repulsive $\phi_{LR}$ part of $\phi$ separately for $\xi\!=\!0.5$ and same values of $\alpha,A$. Representative morphologies are shown for $\xi\!=\!$ (b) $0.8$, (c) $0.7$, (d) $0.6$, and (e) $0.5$, respectively. While strong repulsion results in highly anisotropic string-like clusters, the morphology changes continuously to more compact shapes as attraction wins over. Further dominance of attraction leads to spanning clusters or {\it gels}. Only a part ($1/36^{th}$) of the simulation box is shown for visual clarity.
}
\label{intNmorf}
\end{figure*}
We consider a $2D$ system of interacting mono-disperse particles at fixed density $\rho$ and temperature $T$. The particles interact pair-wise via a short-range attraction, modelled by generalised Lennard-Jones potential \cite{2n-n}, 
\begin{equation}
\phi_{SA}\!=\!4\epsilon[(\sigma/r)^{2\alpha}-(\sigma/r)^\alpha].
\end{equation} 
The length and energy scales of the problem are set by $\sigma$ and $\epsilon$ respectively. The range of attraction can be tuned from a few times of $\sigma$ to an arbitrary small fraction of it by increasing $\alpha$. For example, the range is $2.5\sigma$ and $0.2\sigma$ for $\alpha\!=\!6$ and $\alpha\!=\!18$, respectively. However, the thermodynamics of the system does not change for $\alpha\!\geq\!18$ \cite{2n-n2}. The long-range repulsion experienced by the particles from their surrounding media is considered implicitly by two-body Yukawa potential, 
 \begin{equation}
 \phi_{LR}\!=\!(A\sigma/r)\exp(-r/\xi)
 \end{equation}
The strength of repulsion $A$ and the screening length $\xi$ is expressed in units of $\epsilon$ and $\sigma$, respectively. The effective centro-symmetric potential $\phi\!=\!\phi_{SA}+\phi_{LR}$ (Fig.\ref{intNmorf}(a) {\it inset}) bears an attractive minimum in between a steep hard-core like repulsion and a finite positive repulsive barrier decaying exponentially to zero at large $r$. $\{\alpha, A, \xi\}$ are then three independent parameters that tune $\phi$.  The variations of $\phi$ is shown in Fig.\ref{intNmorf}(a) as a function of $\xi$ for fixed values of $A\!=\!4.0$ and $\alpha\!=\!18$. We note that the global attractive minimum for $\xi\!=\!0.5$ becomes a local minimum and comparable with long-range repulsive minimum for $\xi\!=\!0.8$. Varying $A$ for a fixed value of $\xi$ and suitably chosen $\alpha$ yield similar variations of $\phi$. The results presented next are for $\alpha\!=\!18$ unless otherwise stated. This choice of demonstration is, albeit arbitrary, a reasonable approximation for depletion attraction in globular proteins \cite{gp1,gp2,gp3} and some polymer-coated colloids \cite{colloids1,colloids2,colloids3}.

A system of fixed density $\rho\!=\!0.4$ is used with $56000$ particles in a $376.0\sigma\!\times\!372.0\sigma$ box with periodic boundary conditions along all directions. Time is measured in units of $\tau\!=\!\sqrt{\sigma^2/\epsilon}$ for unit mass. Starting from a random configuration of particles equilibrated at high temperature $T_i\!=\!1.0$, particle trajectories are generated using molecular dynamics (implemented by LAMMPS\cite{lammps}) as the system temperature is linearly ramped down to $T_f\!=\!0.05$ over a period of time $10^4\tau$. Temperature is measured in $\epsilon$ units and maintained by a Langevin thermostat \cite{thermostat}. The equation of motion for the $i$-th particle with position vector ${\bf r}_i$ reads as,
\begin{equation}
\ddot{{\bf r}}_i = -\sum_{j\ne i} \nabla \phi(r)-\nu\dot{{\bf r}}_i +{\bf \zeta}_i
\end{equation}
considering the force due to interaction $\phi(r)$ and frictional drag $\nu\dot{{\bf r}}_i $ from implicit media experienced by the particle. ${\bf \zeta}_i$ is a random force with zero mean and Gaussian variance, $\langle \zeta_i(t_0)\zeta_j(t+t_0)\rangle = 2k_B (T/\nu)\delta_{i,j}\delta(t)$. Boltzmann constant, $k_B$, is chosen to be unity throughout the calculations. Numerical integration of the equation of motion is performed by using velocity Verlet algorithm with time steps $\delta t\!=\!10^{-3}\tau$. We specify that our choice of $T_f$ and $\rho$ is lower than the critical values $T_c\!=\!0.18\pm0.01$ and $\rho_c\!=\!0.6\pm0.1$ for purely attractive systems \cite{critical}. Aggregation sets in during cooling as soon as the system temperature goes below $T_c$. The linear cooling protocol adapted in this work is very slow compared to the typical diffusion timescale ($\sim\!\tau$) of the system and is in contrast to the quench protocol traditionally followed in experiments. With $T_f\!=0.05$ (standard deviation $\sim\!10^{-4}$), we analyse only the part of the trajectories where the number of aggregates does not change over the observation period ($10^3\tau$). The potential energy, however, continues to decay non-exponentially over this observation time. Quantification of such non-ergodic behaviour and other anomalous features of the microscopic dynamics of the system are described elsewhere \cite{2Ddyn} in detail. Here, our main focus is on the local structural properties of the system and its variation with competing interactions. Highly anisotropic non-compact clusters of lateral width $\sigma$ (Fig.\ref{intNmorf}(b)) are observed for strong repulsion over attraction (large $\{A, \xi\}$). As a function of increasing attraction, with decreasing $A$ and/or $\xi$, cluster width increases by sidewise aggregation of particles and compact crystalline islands appear in increasing sizes (Fig.\ref{intNmorf}(c)-(e)). For even higher attraction and near negligible repulsion (small $\{A, \xi\}$), the system consists of one large spanning cluster and several tiny ones. Such states are often referred to as {\it gels} \cite{gel}.

\section{Structural characterisation}
\subsection{Radial and angular correlations}
\begin{figure}[h!]
\centering
\includegraphics[width=0.9\linewidth]{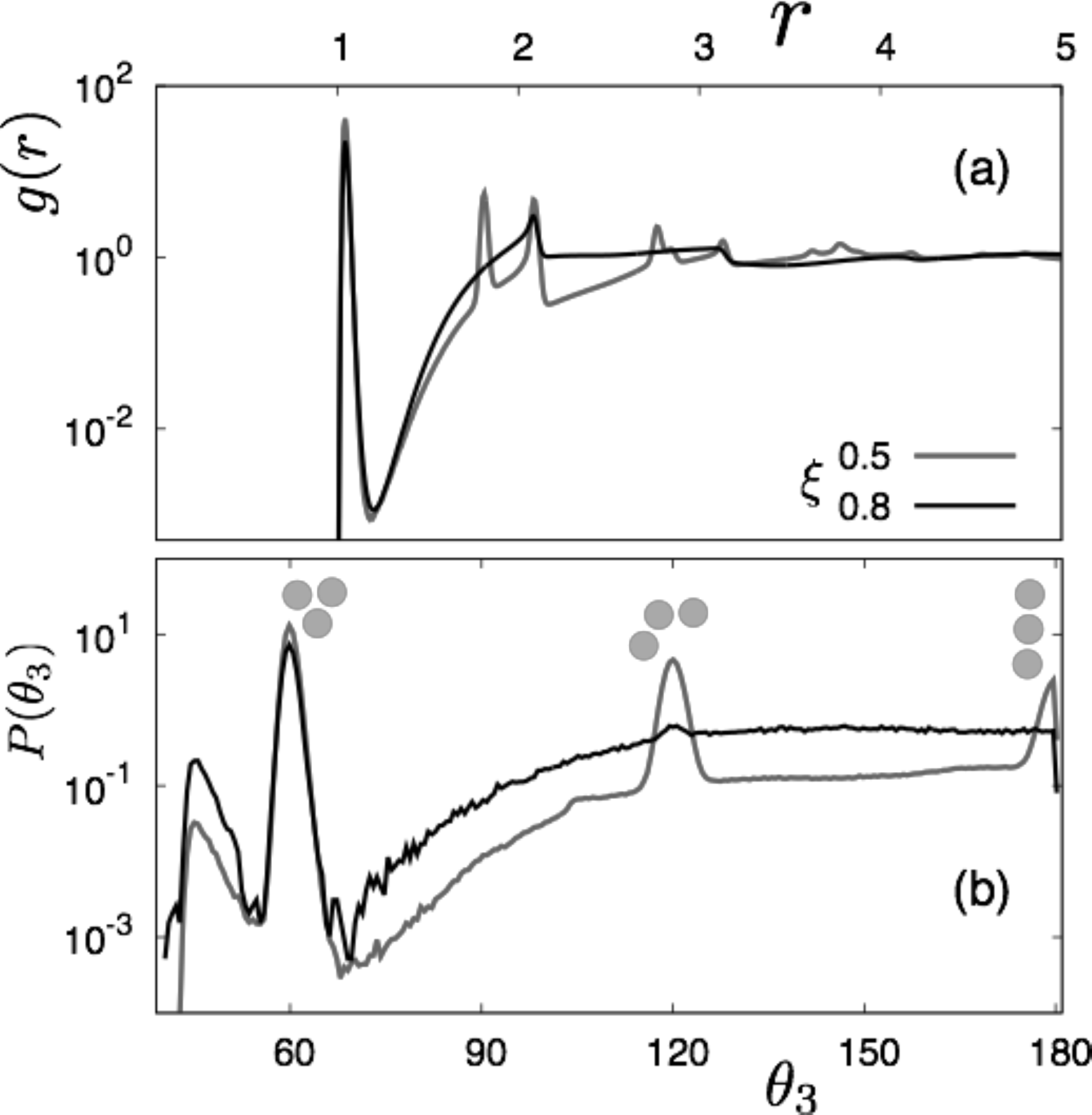}
\caption{
(color online) (a) Radial distribution $g(r)$ of particles for the same configurations shown in Fig.\ref{intNmorf}(b) and (e). While compact structures ($\xi\!=\!0.5$) show symmetric splitting of second peak characteristic of crystalline order in the system, this feature becomes considerably less and anisotropic for non-compact aggregates ($\xi\!=\!0.8$). (b) The distribution of three-body angle $P(\theta_3)$ is computed for the same systems as above. In accord with previous observations, $P(\theta_3)$ shows three preferred peaks at $\pi/3, 2\pi/3$ and $\pi$ for compact crystalline arrangement. Illustrations over each peak are representative of particle arrangements for the three different cases. Last two peaks ($\theta_3\!=\!2\pi/3,\pi$) become less prominent for non-compact structures where particles arrange in a variety of angles to each other, for example, see Fig.\ref{intNmorf}(a).
}
\label{pairangle}
\end{figure}
To characterise the diverse morphologies observed, we now focus on the microscopic features of particle arrangements in simulated configurations. The radial distribution function \cite{LiquidBook},
\begin{equation}
g(r)\!=\!1/\rho\langle\sum\delta(r-r_i)\rangle
\end{equation}
computes the probability of finding an $i$-th particle from an arbitrary central one as a function of the separation between them. Whereas the first peak in $g(r)$ (Fig.\ref{pairangle}(a)) accounts for the nearest neighbours, splitting of the second peak indicates two preferential next nearest neighbour distances, typical of crystalline (compact) arrangements (Fig.\ref{intNmorf}(e)). For non-compact aggregates, the peak heights reduce and the second peak becomes anisotropic. The long shoulder between the preferential positions of first and second peaks accounts for the possible non-crystalline arrangement of particles observed in Fig.\ref{intNmorf}(b). $g(r)$ for other intermediate structures (Fig.\ref{intNmorf}(c)-(d)) naturally falls within these two limits with varied peak heights and anisotropy accordingly. Angular arrangement of particles, being inaccessible by $g(r)$, is studied using three-body angles $\theta_3$. All triads of particles are considered where each particle is the nearest neighbour of at least one other particle and the distribution of three-body angles $P(\theta_3)$ (Fig.\ref{pairangle}(b)) is computed. Three preferential values, $\pi/3, 2\pi/3$ and $\pi$, assumed by $\theta_3$ are indicative of hexagonal symmetry expected in $2D$. Whereas compact aggregates agree with this, non-compact ones naturally deviate as the hexagonal symmetry breaks down and more angular arrangements become possible. The radial and angular features of all percolated configurations are similar to the compact ones, though enhanced as expected.

\subsection{Shape and size statistics}
\begin{figure}[h!]
\centering
\includegraphics[width=\linewidth]{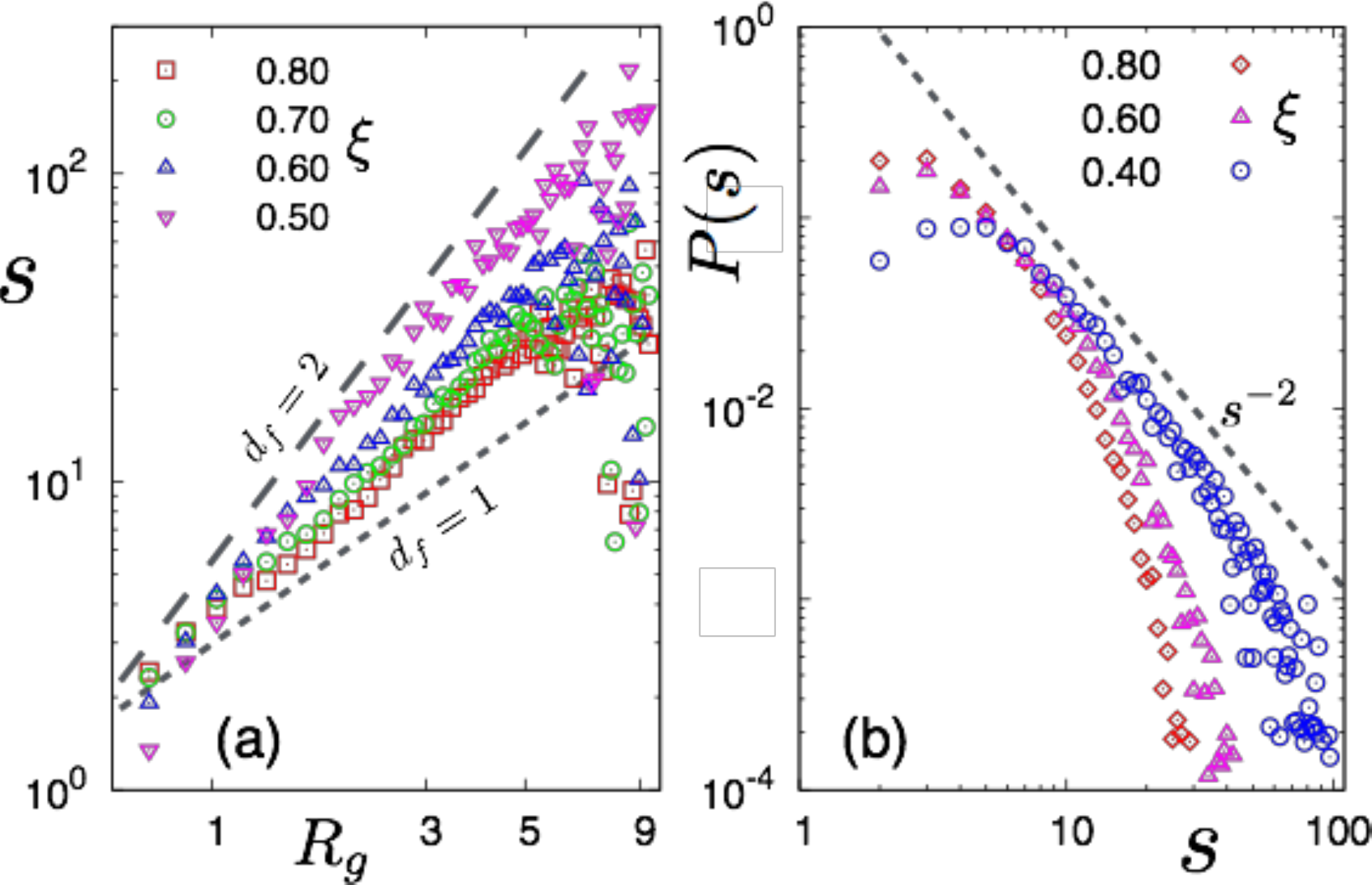}
\caption{
(color online) (a) Fractal dimension $d_f$ of aggregates is computed from the slope of their size $s$ plotted against radius of gyration $R_g$. Here, we plot the values computed for the representative configurations shown in Fig.\ref{intNmorf}(b)-(e). Long- and short-dashed lines have slopes $2$ and $1$ respectively. (b) Size distribution $P(s)$ of aggregates for different values of $\xi$ at $A\!=\!4.0$. Power-law behaviour of $P(s)$ is indicative of percolation at $\xi\!=\!0.4$ in contrast with the exponential behaviour for other values. A dashed line of slope $-2$ is shown for comparison. 
}
\label{shapeNsize}
\end{figure}
Visually evident structural transition of finite-size aggregates is further characterised by their fractal dimension $d_f$. Exploiting the relation between the size $s$ of a cluster and its radius of gyration $R_g$,
\begin{eqnarray}
s&\!\sim\! &R_g^{d_f} \\
R_g^2&\!=\! &1/2s\sum(r_i-r_j)^2 \nonumber
\end{eqnarray}
$d_f$ is obtained for different conformations. For the non-compact (almost linear) and compact (locally crystalline) aggregates, $d_f$ is bounded between $1$ and $2$ corresponding to linear and planar shapes respectively (Fig.\ref{shapeNsize}(a)). As the transition occurs seamlessly with varying competition, it is difficult to define a sharp boundary of structural transition in terms of $d_f$. We consider all configurations with $d_f\!\geq\!1.6$ as compact and the rest as non-compact. All these finite-size aggregates appear in exponentially distributed sizes within the system. These clusters aggregate further as attraction dominates over repulsion and span the system as a single percolating cluster. The size distribution $P(s)$ for the ensemble of clusters thus changes from exponential to algebraic and becomes scale-free \cite{percolation}:
\begin{equation}
P(s)\!\sim\!s^{-\nu} \exp(s/s_0)
\end{equation}
where $\nu$ is the Fisher exponent and exponential part is due to the finite-size effect (Fig.\ref{shapeNsize}(b)). Fitting the cluster size distribution with the above mentioned form, we numerically estimate the Fisher exponent $\nu\!=\!1.98\pm0.04$. This value is very close to the random percolation (RP) model Fisher exponent, $187/91$, in $2D$. Within RP model, $\nu\!<\!2$ is spurious as it implies unphysical divergence of mean cluster size. However, $\nu\!=\!1.91\pm0.06$ is reported experimentally \cite{perco1} and can be explained through a simple extension \cite{perco2} of RP model. Identification and characterisation of the exact nature of this geometric transition require further in-depth investigation and we leave that for future correspondence. We mention that the system density ($\rho\!=\!0.4$) is lower than the typical critical value, $\rho_c\!=\!0.5$, for the RP model. The exponential cut-off $s_0$ is only very weakly dependent on repulsion controlled by $\{A,\xi\}$ and increases with density. Keeping the interaction intact, similar percolation has also been observed as a function of increasing density. The non-compact aggregates percolate at $\rho\!=\!0.5$, consistent with typical random percolation.

\section{Single descriptor for aggregates' morphology phase diagram}
We now compute an effective hard-core diameter \cite{LiquidTheory,LiquidTheory2},
\begin{equation}
\sigma_1\!=\!\int_0^\infty[1-\exp(-\beta\phi_h)]dr
\end{equation}
set by $\phi_{SA}$ where $\phi_h= 4\epsilon(\sigma/r)^{2\alpha}$. $\sigma_1$ is thus a function of $\epsilon$ and $\alpha$. Two control parameters of repulsion, namely, $A$ and $\xi$ can be similarly encoded into another length scale,
\begin{equation}
\sigma_2\!=\!\int_0^\infty[1-\exp(-\beta\phi_{LR})]dr
\end{equation}
where $\beta\!=\!1/(k_BT)$ with Boltzmann constant, $k_B\!=\!1$. As the definitions are not specific to the functional form of the potential, they can be computed for other forms \cite{dlvo,gLJY1,gLJY2,gLJY3} of competing interactions with proper care. For example, it is enough to compute only $\sigma_2$ for {\it 2-Yukawa} \cite{2Y} or {\it Coulomb-Yukawa} \cite{CY} models as the extent of hard-disk interaction is fixed in such models to match the real physical situations at hand.
\begin{figure}[h!]
\centering
\includegraphics[width=\linewidth]{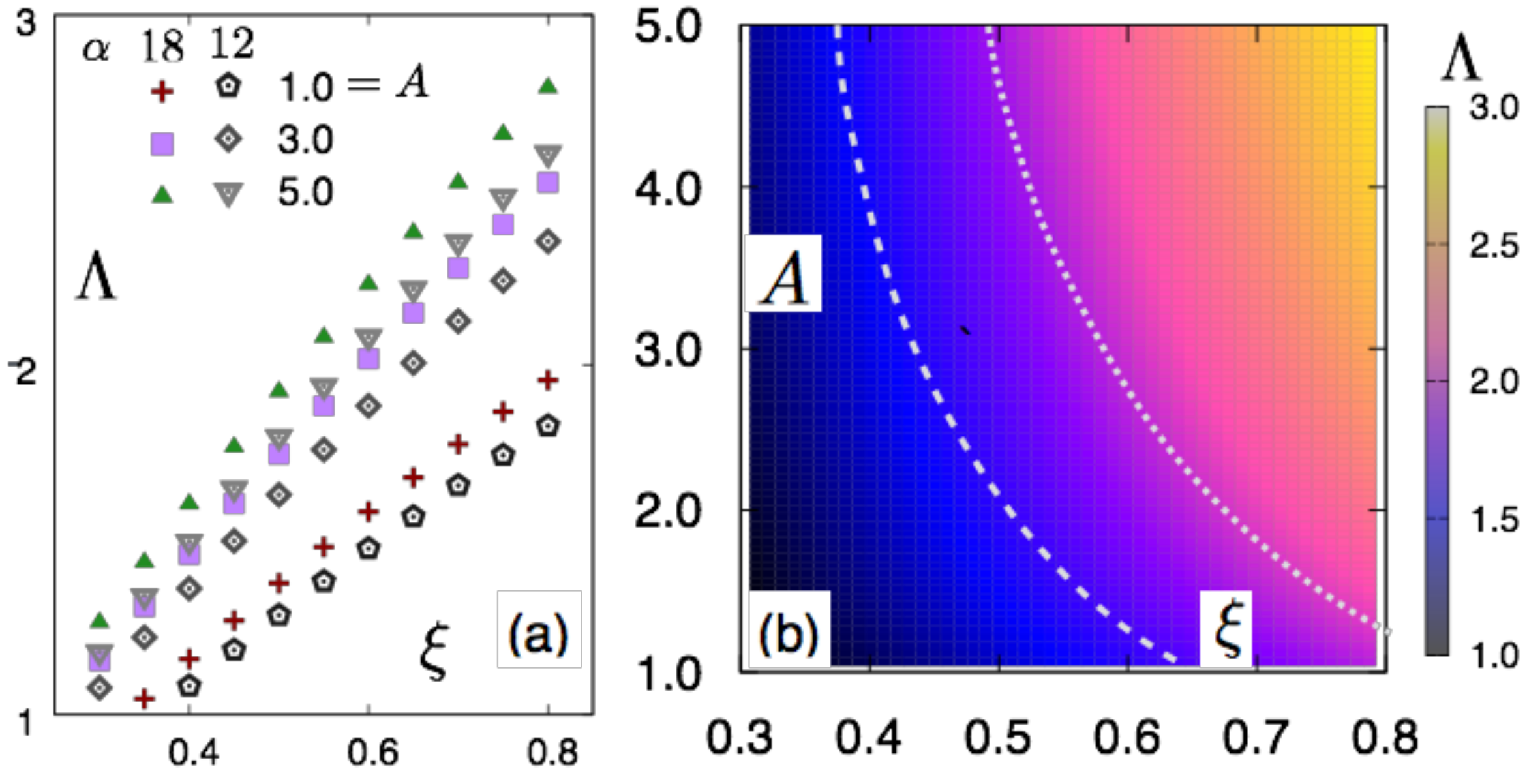}
\caption{
(color online) (a) The dimensionless parameter $\Lambda$ (defined in text) is plotted against $\xi$ for the same set of $A$'s with two different $\alpha\!=\!18$ (open symbol) and $12$ (filled symbols). Please note that one specific value of $\Lambda$ can be obtained through several combinations of $\{\alpha, A, \xi\}$. (b) {\it Morphology phase diagram}: boundaries for different structures drawn on the map of $\Lambda$ in $A$-$\xi$ plane for $\alpha\!=\!18$. Non-compact to compact transition determined by the fractal dimension of respective aggregates is marked by the {\it dotted} line; the {\it dashed} line denotes percolation transition determined from the size statistics of clusters.
}
\label{lmdaNphsdia}
\end{figure}
The effective lengths, $\sigma_1$ and $\sigma_2$, project a useful illustration of the system : consider a system of hard particles with effective diameter $\sigma_1$, each having a concentric soft shell of influence circle with diameter $\sigma_2$.  Inclusion of particles within $\sigma_2$ is possible which in turn changes the perimeter of influence zone of the combined particle system in a non-trivial way thus determining the aggregate morphology. The ratio, $\Lambda\!=\!\sigma_2/\sigma_1$, then encodes all three control parameters, $\alpha$, $A$ and $\xi$ and several combinations of these parameters result in the same $\Lambda$. Fig.\ref{lmdaNphsdia}(a) points out that small $\alpha$-large $A$ is equivalent to large $\alpha$-small $A$ for a range of $\xi$ ($\Lambda$ for $\alpha\!=\!18$, $A\!=\!3.0$ almost coincides with $\alpha\!=\!12$, $A\!=\!5.0$). Larger $\alpha (>\!18)$ will populate the upper left half of Fig.\ref{lmdaNphsdia}(a), whereas smaller $\alpha (<\!12)$ will belong to the lower right half of the same for the specified range of $A$ and $\xi$. The hard disk limit with strict exclusion region \cite{hrdliq,hrdliq2} is recovered for $\Lambda\!=\!1$. Mapping of $\{\alpha,A,\xi\}$ space to a single parameter $\Lambda$ indicates that the morphologies controlled by the former parameters can possibly be represented by the latter. Now, we draw the boundaries of different conformations on a $\Lambda$-map of $\{A,\xi\}$ parameter space for fixed $\alpha\!=\!18$ (Fig.\ref{lmdaNphsdia}(b)). This will serve as the morphology phase diagram of the present model system. Following the appearance of power-law size distribution of clusters, we find that percolated structures or {\it gels} are always formed for parameters giving $\Lambda\!<\!1.5$. For larger $\Lambda$, the aggregates are finite sized and change from compact to non-compact shape around $\Lambda\!\sim\!2.0$ which can be verified from the fractal dimension $d_f$ of respective aggregates.. Taken together with our previous observation, we suggest that lowering $\alpha$ will lead to compact and gel structures whereas non-compact ones are favoured by large $\alpha$ for this system.

\section{Structural randomness and mean energy}
\begin{figure}[h!]
\centering
\includegraphics[width=0.8\linewidth]{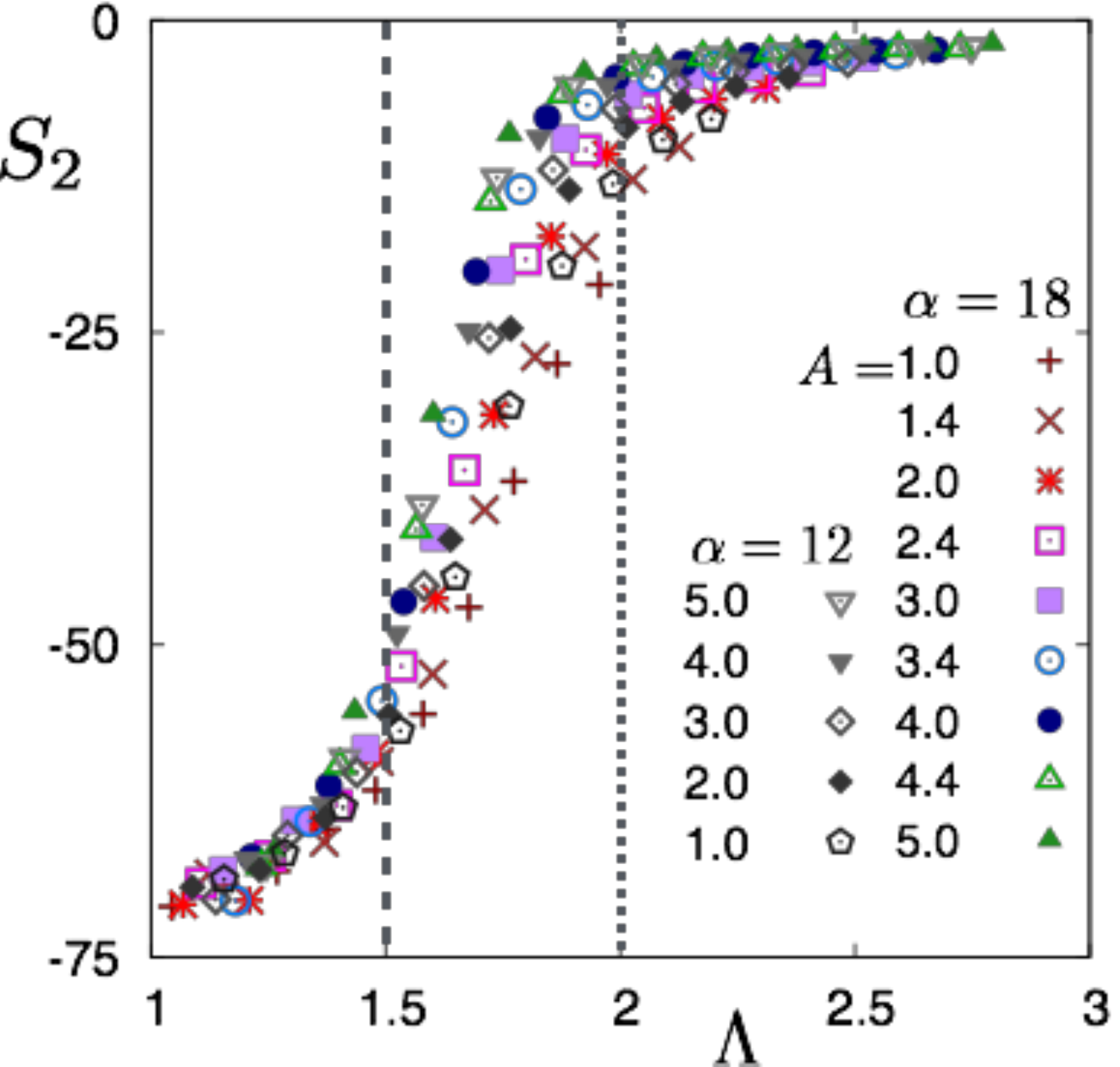}
\caption{
(color online) Two-body excess entropy $S_2$ provides a quantification of positional information. Collapse of $S_2$ with $\Lambda$ offers a unified description of aggregate morphologies independent of their control parameters $\{\alpha,A,\xi\}$. Comparing with morphology phase diagram (Fig.\ref{lmdaNphsdia}(b)) for $\alpha\!=\!18$, we can now identify non-compact ($\Lambda\!>\!2.0$), compact ($1.5\!<\!\Lambda\!<\!2.0$) aggregates and {\it gels} ($\Lambda\!<\!1.5$). The {\it dashed} and {\it dotted} lines are phase boundaries used in Fig.\ref{lmdaNphsdia}(b). As shown, this mapping also holds for $\alpha\!=\!12$. 
}
\label{s2}
\end{figure}
How are the observed morphologies related to their energies? Answering this question requires further quantification of particle arrangements. Positional information of particles can be appropriately expressed as multi-particle expansion of the total excess entropy \cite{exS1} (relative to an ideal gas of same density) truncated at the two-body term,
\begin{equation}
S_2\!=\!-\rho/2\int d{\bf r} \{g(r)\ln g(r)-[g(r)-1]\}.
\end{equation}
This {\it ensemble invariant} measure \cite{exS2} of disorder is $0$ for maximal disorder (ideal gas) and goes to $-\infty$ for the ordered (crystalline) state. When plotted against $\Lambda$ (Fig.\ref{s2}), $S_2$ shows three distinct ranges corresponding to three different morphologies charted out qualitatively in the previous morphology phase diagram (Fig.\ref{lmdaNphsdia}(b)). The non-compact structures for large $\Lambda$ posses higher degrees of randomness close to the random configurations. We note that these structures appear under strong influence of long-range repulsion. One particle can, however, come within the large repulsive influence zone ($\sigma_2\!\ge\!2\sigma_1$) of another particle and is unable to escape due to short-range attraction. This situation evidently favours a collective unidirectional arrangement and opposes the centro-symmetric feature of the effective pair-wise interaction. As the repulsion decreases, particles can not be accommodated within each other's repulsive influence zone and the local hexagonal symmetry favoured in 2D is recovered. $S_2$ reflects this feature by showing a sharp drop for $\Lambda\!<\!2$. As $\sigma_2$ and $\sigma_1$ becomes comparable ($\Lambda\!<\!1.5$) under strong influence of attraction, percolated conformations appear and $S_2$ assumes even larger negative values. Collapse of $S_2$, computed for configurations with $\alpha\!=\!12$ (Fig.\ref{s2}) on the same curve as $\alpha\!=\!18$, establishes the robustness of this mapping against the attraction range. The morphological hierarchy of $2D$ aggregates can then be described in terms of $\Lambda$ alone, and independent of the individual control parameters $\{\alpha,A,\xi\}$.

\begin{figure}[h!]
\centering
\includegraphics[width=0.9\linewidth]{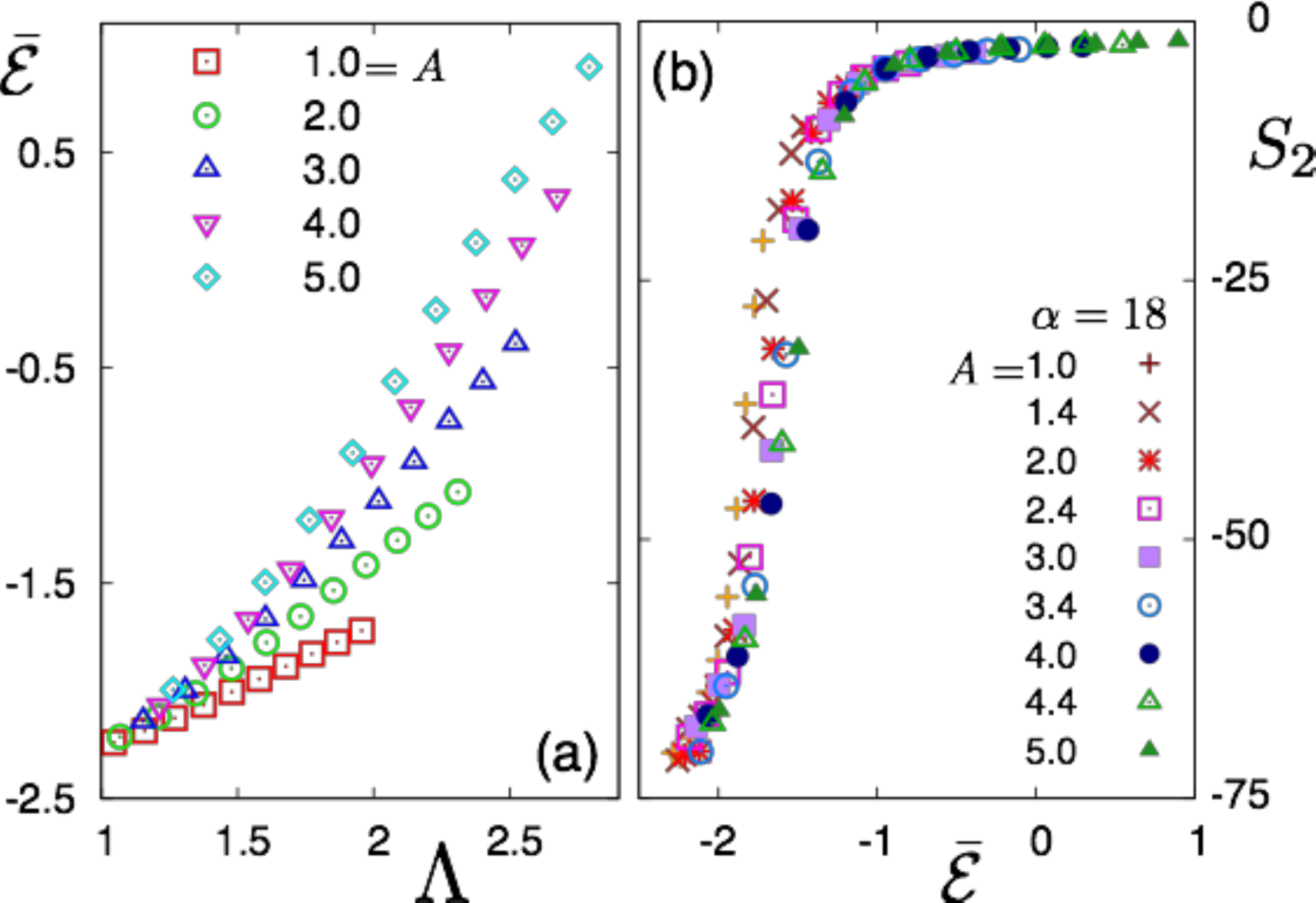}
\caption{
(color online) (a) Average energy $\bar{\mathcal{E}}$ versus $\Lambda$ is linear; the slopes increase with $A$. (b) Plotted against average energy $\bar{\mathcal{E}}$, $S_2$ clearly shows three distinct ranges. These ranges with decreasing energy account for three candidates of morphological hierarchy, non-compact, compact and percolated structures, respectively.
}
\label{elmdas2}
\end{figure}
Since the thermal fluctuations are very small at low temperatures, potential energy plays the dominant role in determining local structures. The average potential energy $\bar{\mathcal{E}}$ is defined as: $\bar{\mathcal{E}}\!=\!(1/N)\sum_i E_i$ where $E_i$ is the potential energy of $i$-th particle due to all other particles present within a cut-off radius of $10\sigma$. $\bar{\mathcal{E}}$ follows a linear relation with $\Lambda$ for fixed $A$ and the slope steepens with increasing $A$ (Fig.\ref{elmdas2}(a)). In other words, as the soft shell of repulsion increases, it is possible to find a set of degenerate $\Lambda$'s with same energy and vice versa. As measured by $S_2$, the morphological hierarchies can now be identified by their distinct ranges of average energy $\bar{\mathcal{E}}$ (Fig.\ref{elmdas2}(b)). Highly disordered non-compact structures ($S_2\!\sim\!0$) are spread over a range of $\bar{\mathcal{E}}$. In contrast, compact structures occurring with a small variance of $\bar{\mathcal{E}}$ show large variation of $S_2$. {\it Gels} have even lower values and smaller spread of both $S_2$ and $\bar{\mathcal{E}}$. These observations certainly call for an in depth understanding which we leave for future investigation.

\section{Discussion}
\begin{figure}[h!]
\centering
\includegraphics[width=0.8\linewidth]{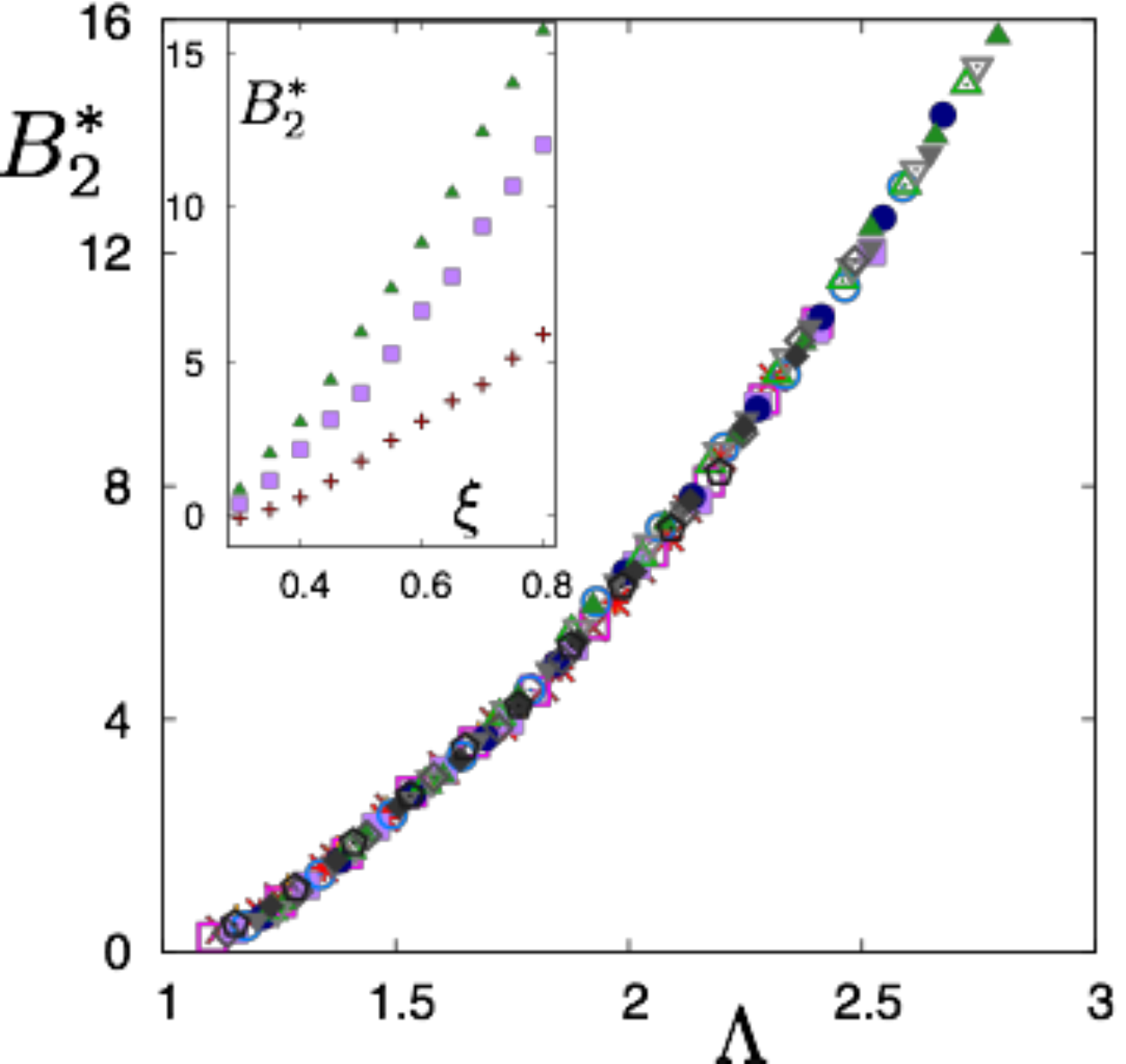}
\caption{
(color online) Second viral coefficient $B_2^*$ computed for the full $\{\alpha,A,\xi\}$ parameter space collapses onto a master curve when plotted against $\Lambda$. Same symbols as in Fig.\ref{s2} have been used. {\it Inset} Please note the spread in $B_2^*$ when plotted as a function of $\xi$ for three different values of $A$ for $\alpha\!=\!18$.
}
\label{B2}
\end{figure}

We note that the knowledge of interactions is necessary for this description. By relating $B_2$, the second virial coefficient, with $\Lambda$, we relax this requirement and connect the description with experiments. Experimental estimation of $B_2$ does not require any prior knowledge of specific potential and/or conformation for a real system. For a simulation study, however, $B_2$ is computed as follows \cite{LiquidBook2},
\begin{equation}
B_2\!=\!-(1/2T)\int d{\bf r} \phi^\prime(r)g(r)
\end{equation}
using both potential and radial distribution. The {\it prime} over $\phi$ denotes the first order spatial derivative of the same. The reduced second virial coefficient, $B_2^*\!=\!2B_2/(\pi\sigma_1)^2$, showing a spread as a function of $A$ and $\xi$, follows a master curve when plotted against $\Lambda$ (Fig.\ref{B2}) for both $\alpha\!=\!18$ and $12$. We note that most experimental efforts specify the overall nature of interactions but do not probe the functional form that corresponds to a given measurement. While the morphologies have been described in several ways, $g(r)$ is, in general, not readily available to calculate $B_2^*$. For example, the existing $B_2^*$ data for globular proteins is inadequate to specify the interactions and/or structures completely. The $B_2^*$-$\Lambda$ master relation provides access to the microscopic length scales from thermodynamic measurements. This collapse has further conceptual implications. $B_2^*$ has already been successfully used \cite{usingB2} to find thermodynamic correspondence among wide range of systems from the van der Waals limit (simple liquids) \cite{vdw,vdw2} to the Baxter limit (sticky hard spheres) \cite{baxter}. This study shows that $B_2^*$ is capable of capturing the morphological correspondence as well and providing a feasible basis for the extension of {\it corresponding states principle} \cite{csp,csp2,csp3} to aggregates.

A formal order parameter approach, mean field \cite{mft,mft2} and beyond \cite{op}, is often challenging for pattern forming systems showing wide variations in local density. Quantification of positional randomness by $S_2$ provides an alternative description capable of relating the local geometry of conformations to a thermodynamic quantity like $\bar{\mathcal{E}}$. Efforts have been made to describe the aggregates employing the integral theory approach \cite{inttheory} for simple liquids. However, such an approach considering the effect of repulsion as a perturbation to attraction is inadequate for the systems presented here where both of the competing interactions are comparable in strength. We have been able to encode the details of competing interactions into two distinct length scales using standard liquid states theory. The procedure is valid over a wide range of control parameters and should also be readily applicable to other model systems as pointed out earlier. Further, these length scales offer an intuitive grip over the physical situation useful for further theoretical developments. The ratio of these length scales, $\Lambda$ can successfully identify all three major candidates, non-compact, compact and percolated structures, forming the morphological hierarchy of aggregates in 2D. We mention that availability of more degrees of freedom in three dimensions may potentially affect the scenario presented here. However, the framework should be valid and we leave the detailed validation for future investigation. Importantly, $\Lambda$ is experimentally accessible through the master relation between $B_2^*$ and $\Lambda$. Given the $S_2$-$\Lambda$ map, the expected morphologies can then be easily predicted. This can be potentially important for the design of new materials with desired functionalities. In conclusion, we have presented a minimal, yet robust description of 2D aggregate morphologies potentially relevant for the self-organisation of diverse nano- and bio-processes. The present work provides the missing information connecting interactions, microscopic structures and thermodynamics. This should motivate further studies on complex fluids probing their structural signatures with higher resolution.

\footnotesize{

}

\end{document}